\documentclass[a4paper]{jpconf}
\usepackage{graphicx}
\usepackage{upgreek}
\RequirePackage{lineno}
\begin{document}
\title{Measurements of open heavy-flavour production with ALICE at the LHC}

\author{Sudipan De$^{1,2}$ for the ALICE collaboration}

\address{1. Departamento de F\'isica Nuclear, Instituto de F\'isica, Universidade de S\~ao Paulo, CEP: 05508-090, S\~ao Paulo, Brasil}
\address{2. Discipline of Physics, School of Basic Sciences, Indian Institute of Technology Indore, Simrol, M.P.- 453552, India}

\ead{Sudipan.De@cern.ch}

\begin{abstract}

In ALICE, open heavy-flavour production is studied through the measurements of the leptons (electrons and muons) from heavy-flavour hadron decays at central and forward rapidity and via the reconstruction of D-meson hadronic decays at  mid-rapidity. An overview of the open heavy-flavour production with ALICE in pp ($\sqrt{s}$ = 2.76 TeV and 7 TeV), p--Pb ($\sqrt {s_{\rm NN}}$ = 5.02 TeV) and Pb--Pb ($\sqrt {s_{\rm NN}}$ = 2.76 TeV) collisions will be presented. We will discuss the production cross sections, modifications of the transverse momentum distributions, azimuthal anisotropic emissions and correlations with hadrons in comparison with various theoretical predictions.     
\end{abstract}

\section{Introduction}
A deconfined state of quarks and gluons, generally known as the Quark--Gluon Plasma (QGP), is expected to be formed in ultrarelativistic heavy-ion collisions. Due to their large mass, heavy-quarks, i.e.\ charm and beauty quarks, are produced at the early stages of the collisions via hard scattering processes. Therefore, they can experience the complete evolution of the system and interact with the created hot medium. Theoretical model calculations based on Quantum Chromodynamics (QCD) predict a mass-dependent energy loss due to the mass-dependent restriction of the phase space of the gluon radiation. Therefore, heavy quarks are expected to lose less energy compared to the light quarks~\cite{mass1,mass2}. In addition, gluons are considered to lose more energy than quarks due to their stronger colour coupling~\cite{mass3}. Experimentally this can be studied by measuring the nuclear modification factor of hadrons containing heavy quarks, which is defined as the ratio between the differential yield measured in Pb--Pb (or p--Pb) collisions and the cross section measured in pp collisions multiplied by the average nuclear overlap function. The degree of the thermalisation of the heavy quarks in the medium can be studied by measuring the anisotropic flow parameter $v_{2}$, second harmonic of the Fourier expansion of the particle azimuthal distribution, of the heavy-flavour particles at low transverse momentum ($p_{\rm T}$). The elliptic flow measurement at high $p_{\rm T}$ gives information about the path-length dependence of energy loss of the heavy quarks. Measurements of heavy-flavour particles in pp collisions are studied to test the perturbative QCD (pQCD) calculations and to set a baseline for both Pb--Pb and p--Pb collisions. Studies of heavy-quark production as a function of charged-particle multiplicity in pp collisions are expected to be sensitive to the interplay between hard and soft QCD processes and therefore they give insight into the multi-parton interactions (MPI) ~\cite{MPI}. Measurements of open heavy-flavour production in p--Pb collisions can be used to disentangle Cold Nuclear Matter (CNM) effects, such as the modification of nuclear parton distribution functions (PDF)~\cite{CNM1, CNM2} and $k_{\rm T}$ broadening, from final-state effects occurring in the QGP. In this article we report the open heavy-flavour production via hadronic decays of D mesons (D$^{0}$, D$^{+}$ and D$^{*+}$) and the semileptonic decays of charm and beauty hadrons in pp collisions at $\sqrt{s}$ = 2.76 TeV and 7 TeV, in p--Pb collisions at $\sqrt {s_{\rm NN}}$ = 5.02 TeV and in Pb--Pb collisions at $\sqrt {s_{\rm NN}}$ = 2.76 TeV. 

The details of the ALICE detector are described in~\cite{ALICE1, ALICE2}. Charged particles were reconstructed and identified in the central barrel, which consists of the Inner Tracking System (ITS), Time Projection Chamber (TPC) and Time-Of-Flight (TOF) detector at mid-rapidity ($|\eta| < 0.9$). Muons were identified using the Muon Spectrometer covering $-4 < \eta < -2.5$. Triggering and multiplicity determination are performed by two arrays of scintillator detectors V0A ($2.8 < \eta < 5.1$) and V0C ($-3.7 < \eta < -1.7$) placed on both sides of the interaction point. ElectroMagnetic Calorimeter (EMCal) is used to select the high $p_{\rm T}$ electrons candidates. D-meson candidates are selected based on particle identification of the decay tracks and the decay topology using ITS, TPC and TOF detectors. The details of the reconstruction procedure can be found in~\cite{Dmeson}. Electron candidates are reconstructed using ITS, TPC, TOF and EMCal detector and muons are reconstructed with the Muon Spectrometer. The detailed procedures to measure the heavy-flavour hadrons decay electrons and muons are described in~\cite{electron} and ~\cite{muon1}, respectively.         

\section{Results}

\subsection{Results in pp collisions}

\begin{figure}[tbp]
\begin{center}
\includegraphics[scale=0.35]{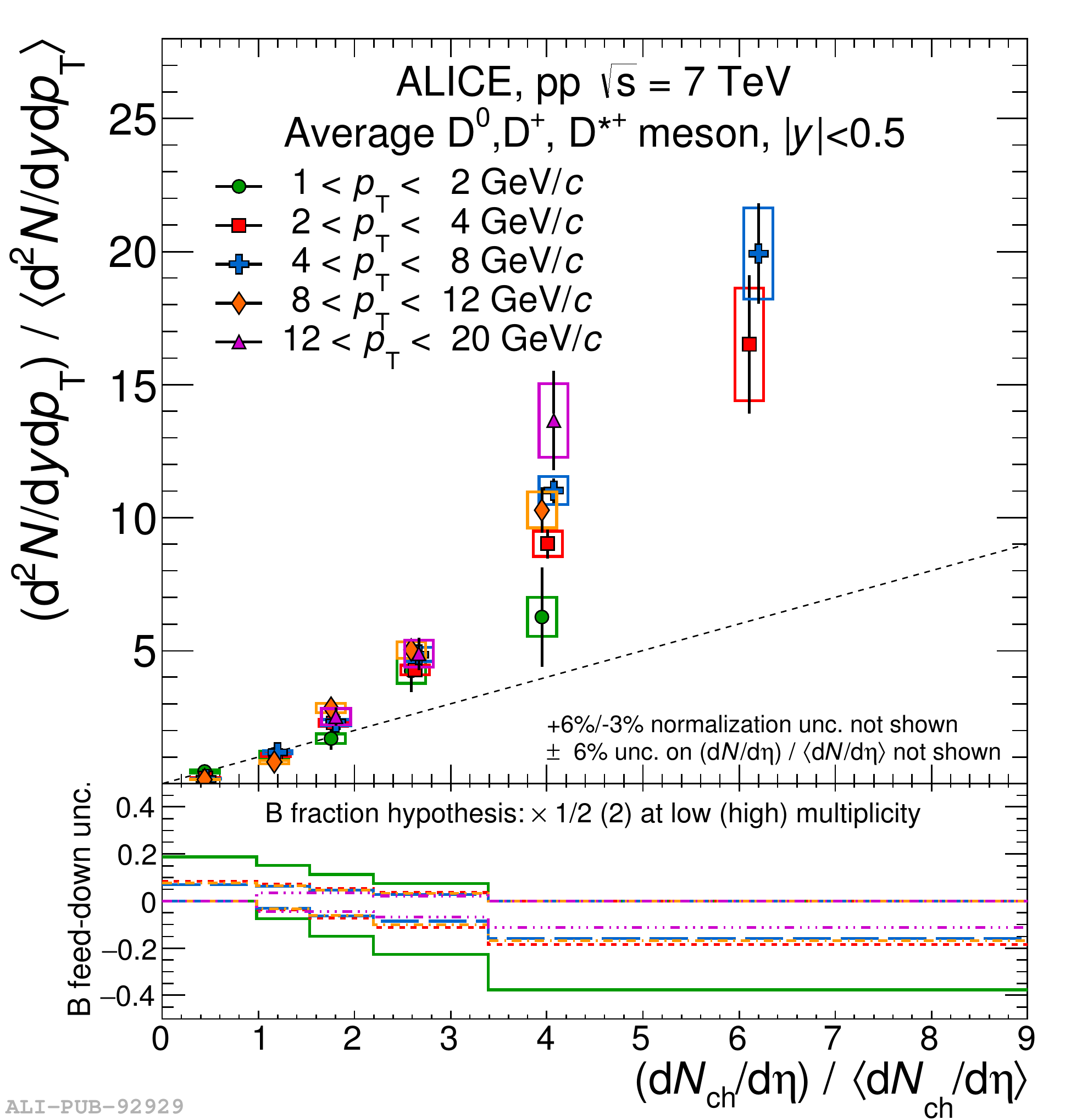}
\includegraphics[scale=0.35]{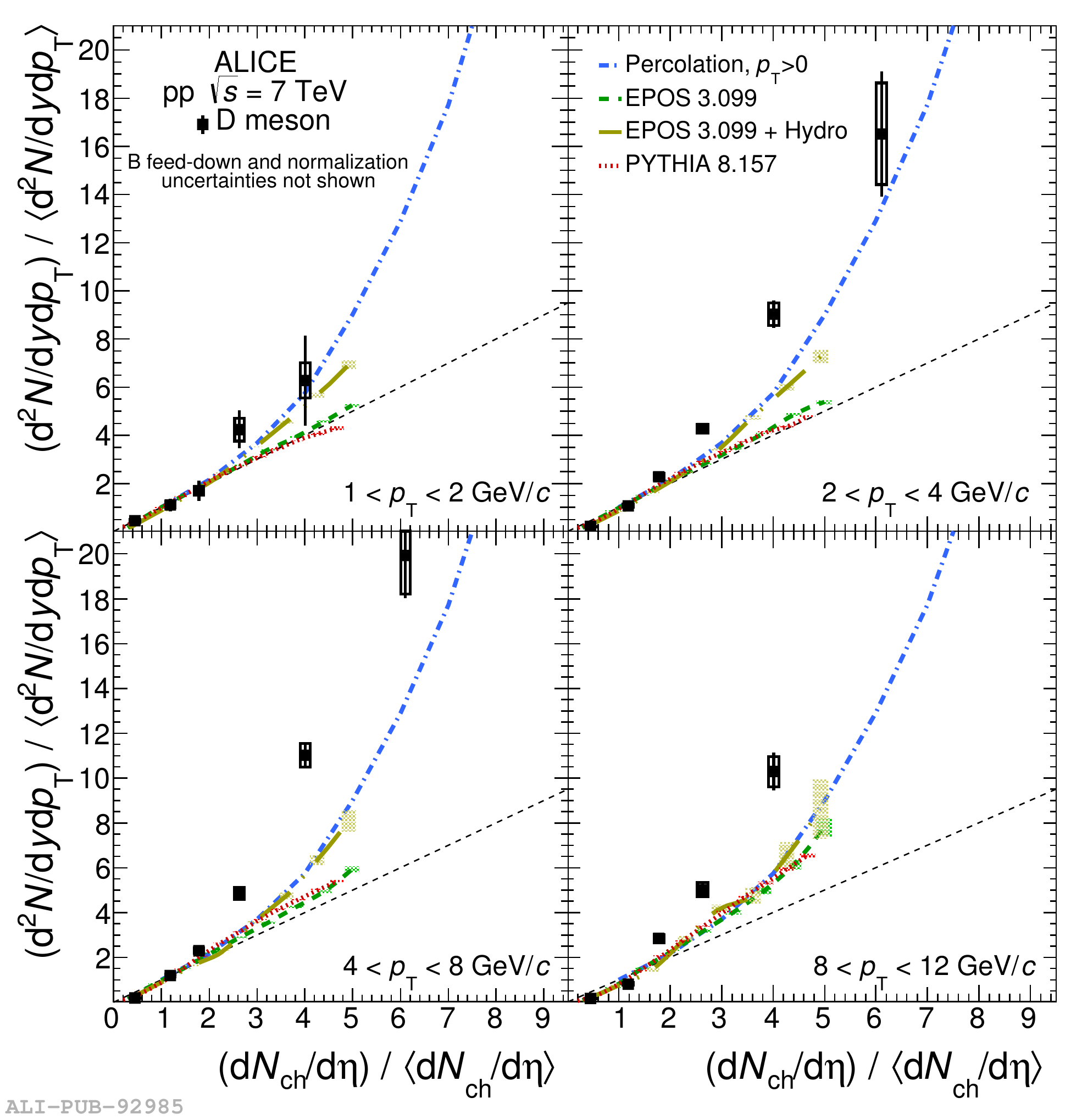}
\caption{
Left: Self-normalised yield of D mesons as a function of charged particle multiplicity in pp collisions at $\sqrt{s}$ = 7 TeV~\cite{Dmeson}. Different colours are for different $p_{\rm T}$ intervals. Right: Self-normalised yield of D mesons as a function of charged-particle multiplicity in different $p_{\rm T}$ intervals with different model predictions~\cite{Dmeson}. Different colour lines show the predictions from different theoretical calculations. Uncertainties on the data points are the statistical uncertainties and boxes represent the systematic uncertainties. The black dotted lines show the linear increase to guide the eye.      
}
\label{fig:pp1}
\end{center}
\end{figure}

The differential cross sections of prompt D mesons (D$^{0}$, D$^{+}$, D$^{*+}$)~\cite{Dmeson1, Dmeson2}, heavy-flavour hadron decay leptons~\cite{electron1, muon1} and beauty-hadron decay electrons~\cite{electronb1, electronb2} were measured in pp collisions at $\sqrt{s}$ = 2.76 TeV and 7 TeV. The results are well described by the pQCD calculations~\cite{FONLL, GM} within the systematic uncertainties. Differential cross section of prompt D$^{0}$ meson was measured down to $p_{\rm T}$ = 0 at $\sqrt{s}$ = 7 TeV and results are well described by the pQCD calculations~\cite{Dmeson3}. The left panel of Fig.~\ref{fig:pp1} shows the self-normalised D-meson yield as a function of charged-particle multiplicity at $\sqrt{s}$ = 7 TeV at mid-rapidity ($|y| <$ 0.5). It shows a stronger than the linear increase as a function of the multiplicity for all the $p_{\rm T}$ intervals. The results in the different $p_{\rm T}$ ranges are found to be in agreement within the uncertainties. The results are found in qualitative agreement with the model predictions, which contain MPI and hydrodynamic effects (right panel of Fig.~\ref{fig:pp1}). It suggests that MPIs play an important role in heavy-flavour production in case of high-multiplicity events in pp collisions~\cite{Dmeson}.        

\subsection{Results in p--Pb collisions}
\begin{figure}[tbp]
\begin{center}
\includegraphics[scale=0.32]{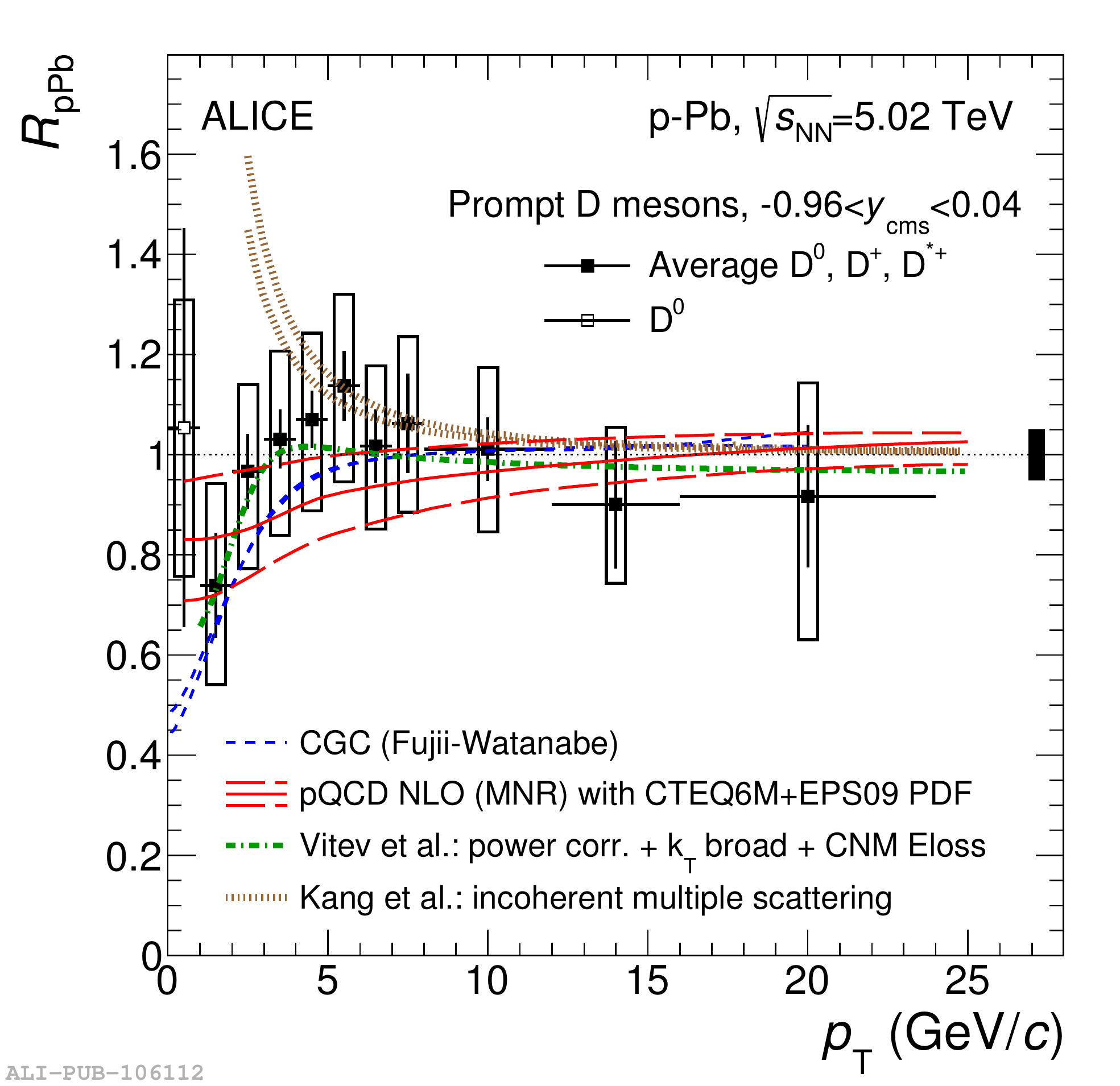}
\includegraphics[scale=0.35]{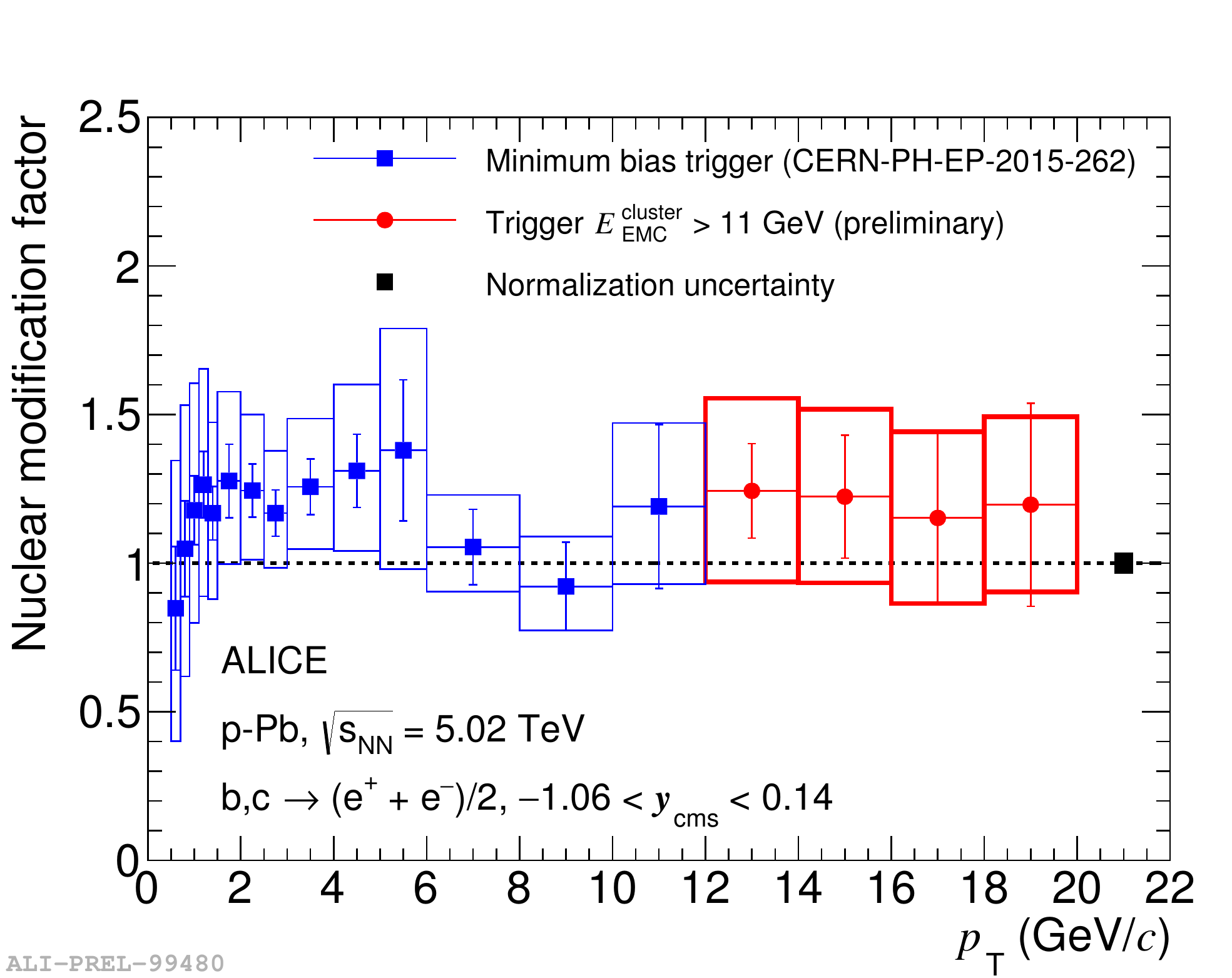}
\caption{
 Left: Nuclear modification factor $R_{\rm pPb}$ of D mesons in p--Pb collisions at $\sqrt {s_{\rm NN}}$ = 5.02 TeV~\cite{Dmeson3}. Different lines represent different model predictions. Right: The $R_{\rm pPb}$ of heavy-flavour hadron decay electrons. Uncertainties on the data points are the statistical uncertainties and boxes represent the systematic uncertainties.          
}
\label{fig:pPb1}
\end{center}
\end{figure}

\begin{figure}[tbp]
\begin{center}
\includegraphics[scale=0.35]{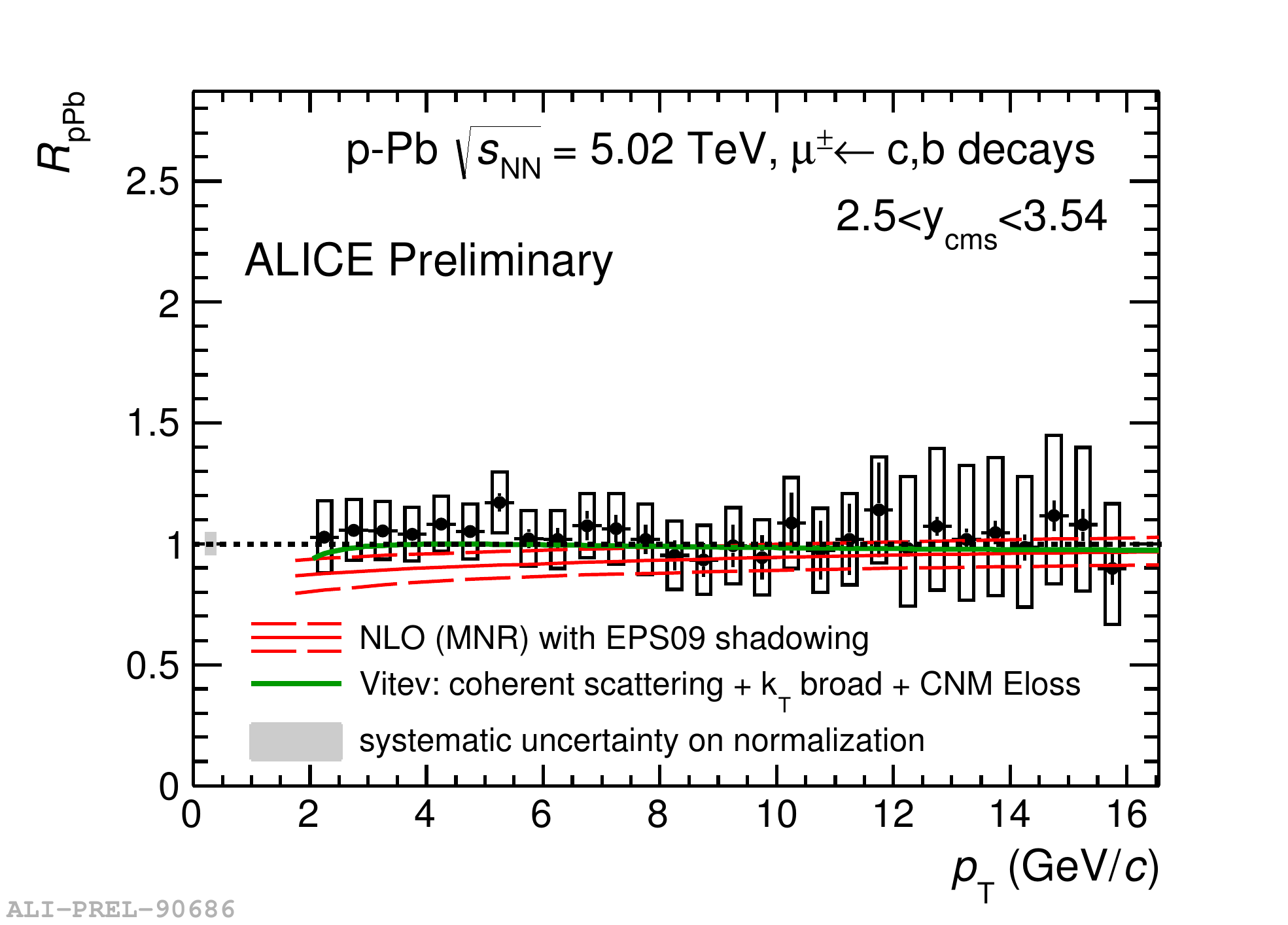}
\includegraphics[scale=0.35]{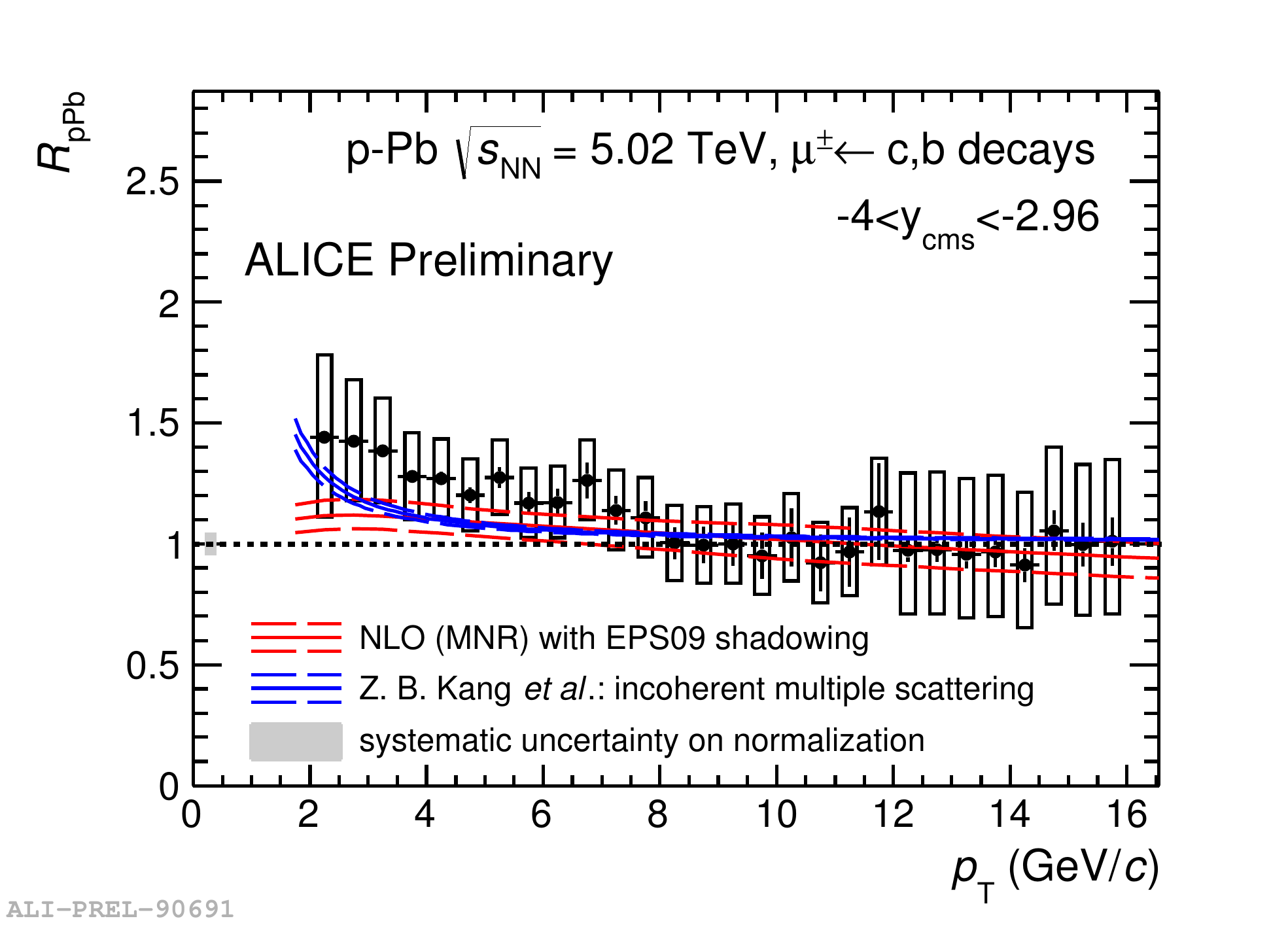}
\caption{
Nuclear modification factor $R_{\rm pPb}$ of heavy-flavour hadrons decay muons in p--Pb collisions at $\sqrt {s_{\rm NN}}$ = 5.02 TeV at forward rapidity (2.5 $< y_{\rm cm} <$ 3.54) (left) and at backward rapidity (-4 $< y_{\rm cm} <$ -2.96) (right). Uncertainties on the data points are the statistical uncertainties and boxes represent the systematic uncertainties. The lines show the different model predictions.          
}
\label{fig:pPb2}
\end{center}
\end{figure}

The left panel of the Fig.~\ref{fig:pPb1} shows the combined measurement of the average nuclear modification factor ($R_{\rm pPb}$) of prompt D$^{0}$, D$^{+}$ and D$^{*+}$ mesons as a function of $p_{\rm T}$ for 1 $< p_{\rm T} <$ 20 GeV/{\it c} and D$^{0}$ mesons for 0 $< p_{\rm T} <$ 1 GeV/{\it c} at mid-rapidity in p--Pb collisions at $\sqrt {s_{\rm NN}}$ = 5.02 TeV~\cite{Dmeson3}. Results are compared with the theoretical results containing CNM effects~\cite{Dmeson3}. All the model predictions describe the data well for the entire $p_{\rm T}$ range within the uncertainties. The right panel of Fig.~\ref{fig:pPb1} shows the $R_{\rm pPb}$ of heavy-flavour hadron decay electrons as a function of $p_{\rm T}$ for 0.5 $< p_{\rm T} <$ 20 GeV/{\it c} at mid-rapidity~\cite{electron}. The $R_{\rm pPb}$ of prompt D mesons and heavy-flavour hadron decay electrons is consistent with unity within uncertainties. Fig.~\ref{fig:pPb2} shows the $R_{\rm pPb}$ of the heavy-flavour hadron decay muons at both forward rapidity (2.5 $< y_{\rm cm} <$ 3.54, proton-beam direction) and backward rapidity ($-4 < y_{\rm cm} < -2.96$, Pb-beam direction). The $R_{\rm pPb}$ is consistent with unity at both at forward and backward rapidity. However there is an enhancement above unity for $2 < p_{\rm T} <$ 4 GeV/{\it c} for the backward rapidity region. In both cases the results are well described by the model predictions within uncertainties. 

\subsection{Results in Pb--Pb collisions}
\begin{figure}[tbp]
\begin{center}
\includegraphics[scale=0.35]{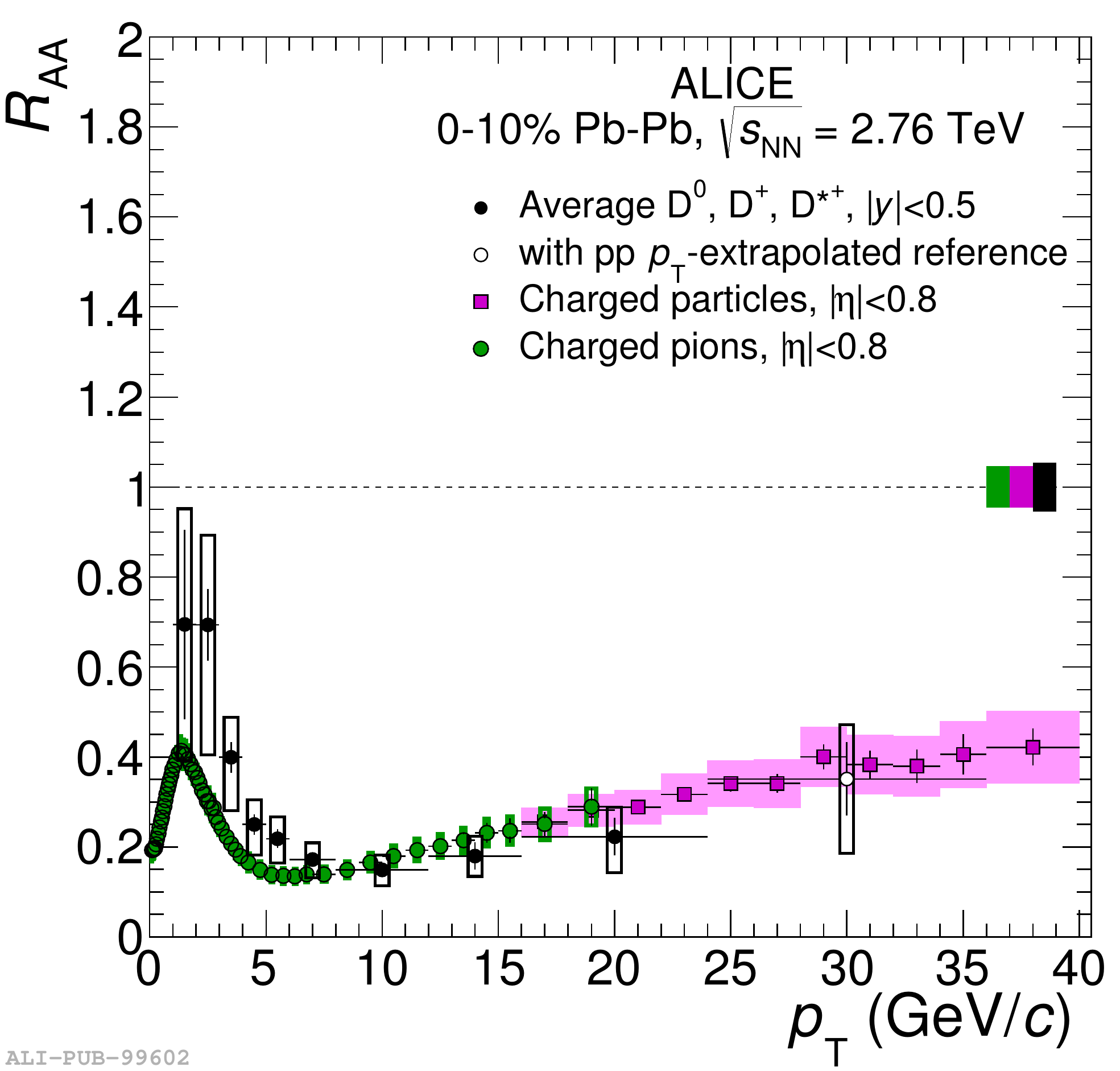}
\includegraphics[scale=0.35]{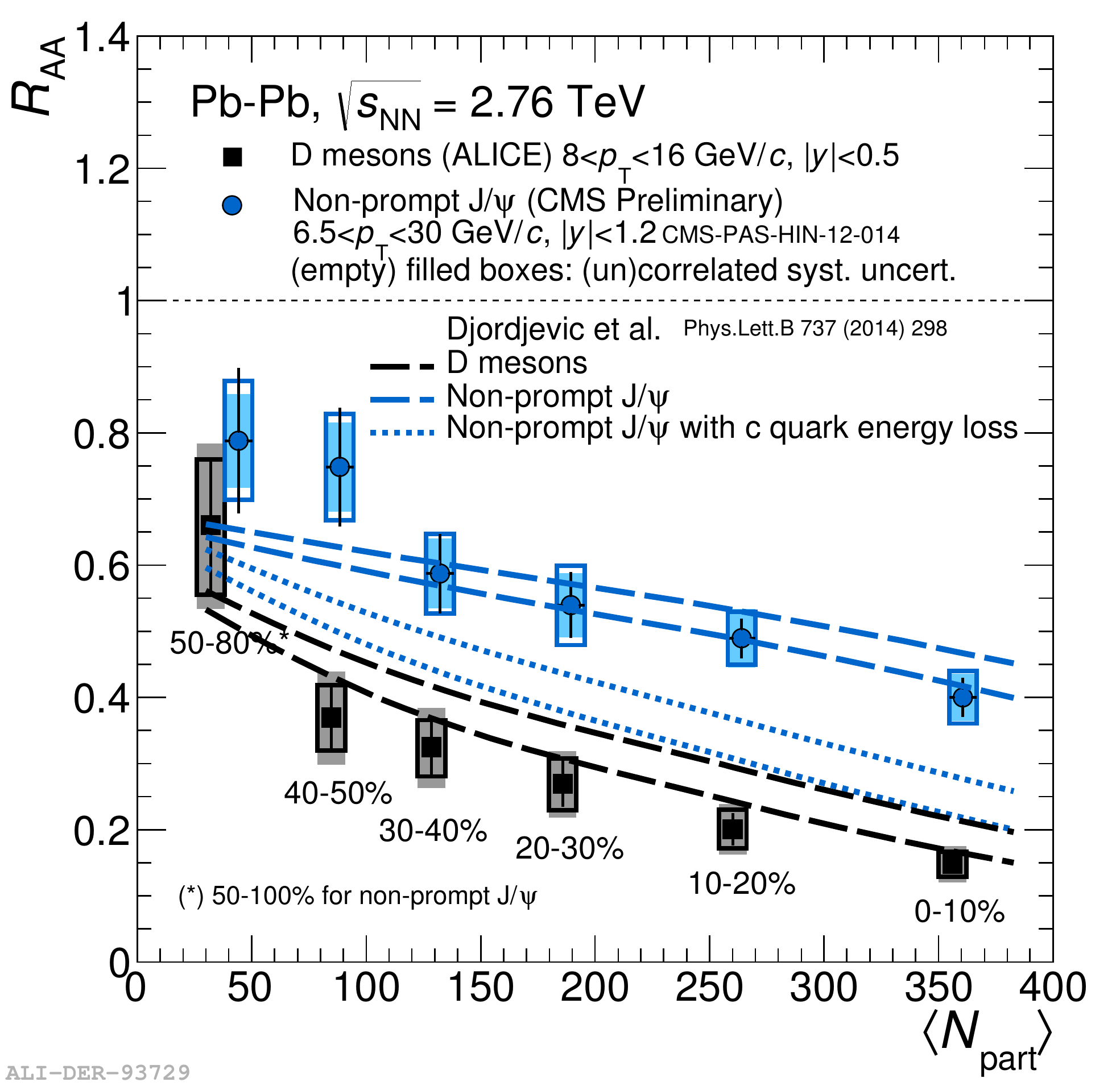}
\caption{
Left: Nuclear modification factor $R_{\rm AA}$ of D mesons (black points) in Pb--Pb collisions with the $R_{\rm AA}$ of charged-particles (magenta) and pions (green) at $\sqrt {s_{\rm NN}}$ = 2.76 TeV for 0--10\% centrality~\cite{pion}. Right: The $R_{\rm AA}$ of D mesons (black points) in comparison with non-prompt J/$\uppsi$ (blue points)~\cite{jsi}. The dotted and dashed lines are from model predictions. Uncertainties on the data points are the statistical uncertainties and boxes represent the systematic uncertainties.          
}
\label{fig:PbPb1}
\end{center}
\end{figure}

The left panel of Fig.~\ref{fig:PbPb1} shows the nuclear modification factor ($R_{\rm AA}$) as a function of $p_{\rm T}$ of D mesons for central Pb--Pb collisions at $\sqrt {s_{\rm NN}}$ = 2.76 TeV at mid-rapidity. A strong suppression of the D-meson yield is observed at intermediate and high $p_{\rm T}$. The D-meson $R_{\rm AA}$ is compatible with the charged particle and pion $R_{\rm AA}$ within the uncertainties~\cite{pion}. The right panel of Fig.~\ref{fig:PbPb1} shows the $R_{\rm AA}$ of prompt D mesons and non-prompt J/$\uppsi$ from beauty-hadron decays from the CMS collaboration~\cite{CMS1,CMS2} as function of the Pb--Pb collision centrality. A larger suppression for D mesons than for non-prompt J/$\uppsi$ is observed in central collisions~\cite{jsi}. The data are well reproduced by the model calculations including mass dependent energy loss~\cite{color2}. 

\begin{figure}[tbp]
\begin{center}
\includegraphics[scale=0.35]{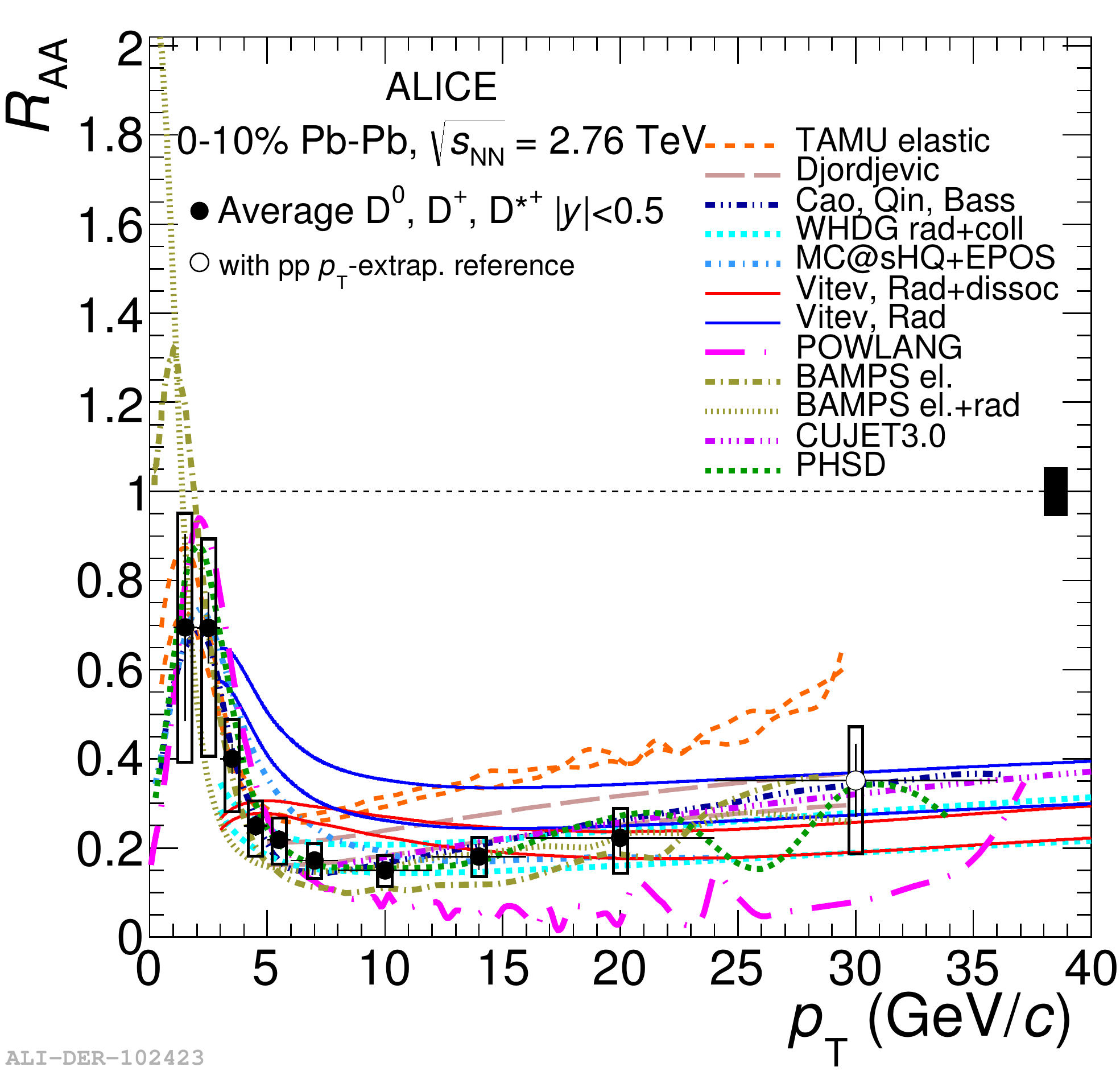}
\includegraphics[scale=0.35]{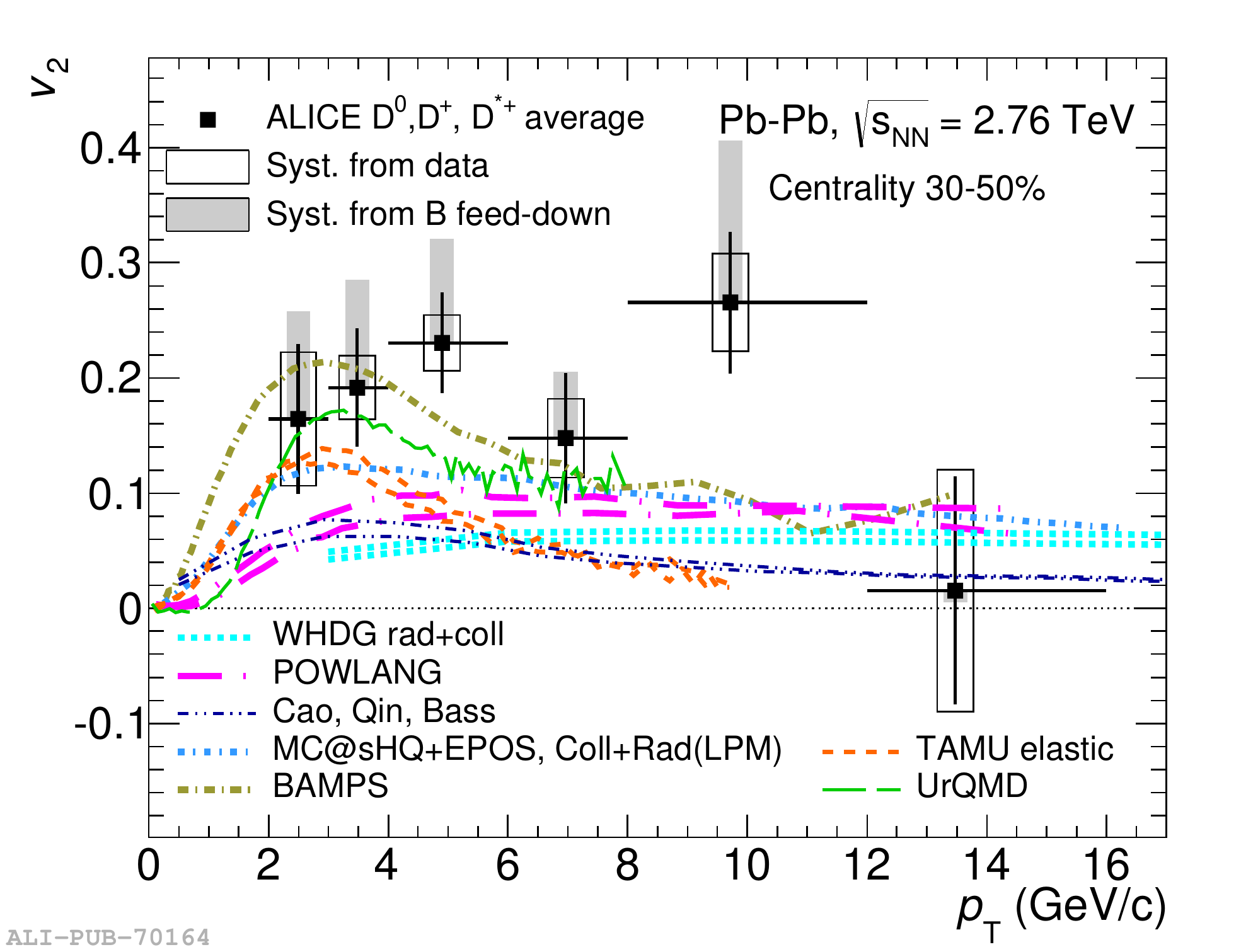}
\caption{
Nuclear modification factor $R_{\rm AA}$ (left) and elliptic flow $v_{2}$ (right) of D mesons in Pb--Pb collisions at $\sqrt {s_{\rm NN}}$ = 2.76 TeV~\cite{pion, PbPb2}. The different lines represent various model predictions. Uncertainties on the data points are the statistical uncertainties and boxes represent the systematic uncertainties.          
}
\label{fig:PbPb2}
\end{center}
\end{figure}       

The left panel of Fig.~\ref{fig:PbPb2} shows the $R_{\rm AA}$ of D mesons in central Pb--Pb collisions (0--10\%) and right panel shows the elliptic flow ($v_{2}$) measurements of D mesons for mid-central collisions (30--50\%) with the different model predictions~\cite{pion, PbPb2}. The positive $v_{2}$ of D mesons at low $p_{\rm T}$ indicates that charm quarks participate in the collective expansion of the medium. The models which explain the strong suppression seen in the $R_{\rm AA}$ tend to underestimate the $v_{2}$. The simultaneous model description of the heavy-flavour $R_{\rm AA}$ and $v_{2}$ is still challenging and can therefore help to constrain the theoretical understanding of the interaction of heavy quarks in the medium.

\section{Summary}
We presented the open heavy-flavour production in pp collisions at $\sqrt{s}$ = 2.76 TeV and 7 TeV, in p--Pb collisions at $\sqrt {s_{\rm NN}}$ = 5.02 TeV and in Pb--Pb collisions at $\sqrt {s_{\rm NN}}$ = 2.76 TeV with ALICE at the LHC. The results from pp collisions are well described by the pQCD calculations within the uncertainties. Measurements of the multiplicity dependent self-normalised yield of D mesons in pp collisions indicate the MPI contribute to high-multiplicity events and are affecting heavy-flavour production. Strong suppression of heavy-flavour yield is observed in most central (0--10\%) Pb--Pb collisions at intermediate and high $p_{\rm T}$. In p--Pb collisions the $R_{\rm pPb}$ is measured at mid-rapidity for D mesons and heavy-flavour hadron decay electrons as well as at forward and backward rapidity for heavy-flavour hadrons decay muons. For all the cases, the results are consistent with unity indicating that the initial state effects are small and the strong suppression of the heavy-flavour yield observed in Pb--Pb collisions is due the presence of the hot and dense QCD matter. The nuclear modification factor of prompt D mesons is smaller than that of non-prompt J/$\uppsi$ suggesting a mass dependent energy loss of the heavy quarks. The positive $v_{2}$ of the D mesons indicates collective motion of the charm quarks in the medium at low $p_{\rm T}$. The theoretical models are challenged by simultaneously reproducing the $R_{\rm AA}$ and $v_{2}$.

\end{document}